\documentclass[conference]{IEEEtran}

\usepackage{amsmath,amsfonts,amsthm}
        
\usepackage{graphicx}
\usepackage[dvipsnames]{xcolor}
\usepackage[english]{babel}
\usepackage{algorithm}
\usepackage{algorithmic}

\newtheorem{theorem}{Theorem}

\newtheorem{corollary}{Corollary}
\newtheorem{assumption}{Assumption}
\newtheorem{proposition}{Proposition}

\IEEEoverridecommandlockouts

\begin{document}

\title{Full Exploitation of Limited Memory in Quantum Entanglement Switching
{\footnotesize }
\thanks{The research work was supported by the Army Research Office MURI under the project number W911NF2110325 and by the National Science Foundation under project numbers EEC-1941583 CQN ERC and CNS 1955744.\\
ISBN 978-3-903176-57-7© 2023 IFIP.}}

\author{\IEEEauthorblockN{Panagiotis~Promponas\IEEEauthorrefmark{1},
V\'ictor Valls\IEEEauthorrefmark{2}, and Leandros Tassiulas\IEEEauthorrefmark{1}\\
\{panagiotis.promponas@yale.edu, victor.valls@ibm.com, leandros.tassiulas@yale.edu\}}
 \IEEEauthorblockA{\IEEEauthorrefmark{1} \textit{Department of Electrical Engineering and Institute for Network Science, Yale University, USA}}
\IEEEauthorblockA{\IEEEauthorrefmark{2} \textit{IBM Research Dublin}}
}



\maketitle

\begin{abstract}

We study the problem of operating a quantum switch with memory constraints. In particular, the switch has to allocate quantum memories to clients to generate link-level entanglements (LLEs), and then use these to serve end-to-end entanglements requests. The paper's main contributions are (i) to characterize the switch's capacity region, and (ii) to propose a memory allocation policy (MEW) that is throughput optimal. The worst-case time complexity of MEW is exponential on the system parameters. However, when the requests are bipartite and the LLE attempts are always successful, we propose a variant of MEW (MEW2) that has polynomial time complexity. We evaluate the proposed policies numerically and illustrate their performance depending on the requests arrivals characteristics and the time available to obtain a memory allocation. 

\end{abstract}

\section{Introduction}

Quantum computing will transform the world by allowing us to solve problems that are too complex for classical computers \cite{preskill2018quantum} (e.g., Shor's algorithm \cite{Sho99}). However, we are still nowhere near that day. Quantum programs of meaningful size require quantum computers with thousands of qubits  \cite{ibmroadmap}, which is far from the number of qubits that quantum computers currently have \cite{ibm127,arute2019quantum}.

\begin{figure}[t!]
\centering
\includegraphics[width=0.9\columnwidth]{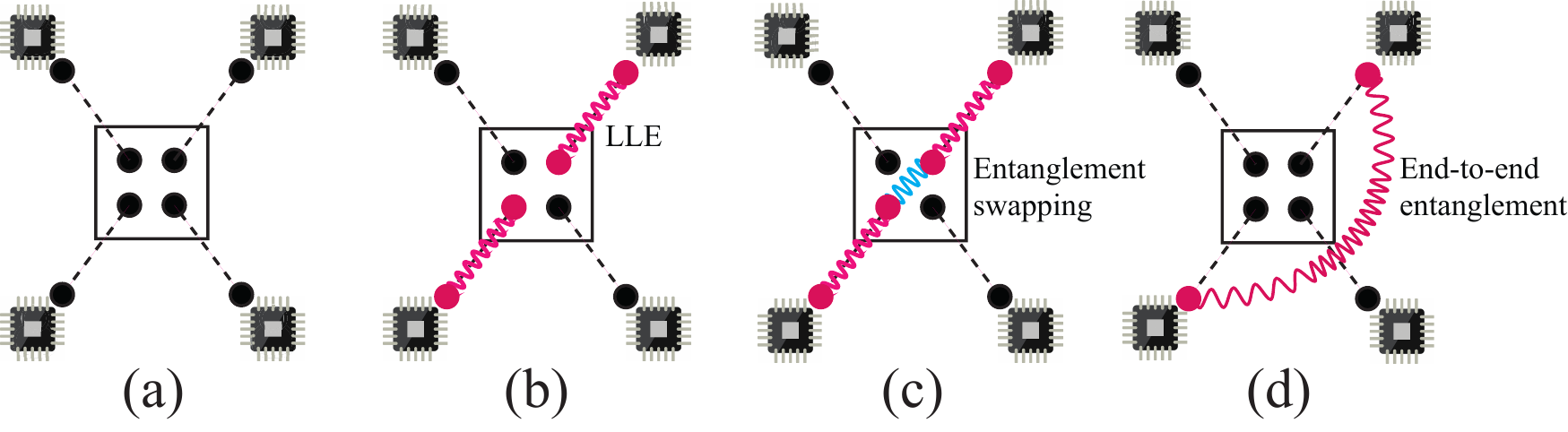}
\caption{Illustrating the operation of a quantum switch. (b) LLEs are created between the switch and the clients (e.g., quantum processors). (c) The switch performs an entanglement swapping operation. (d) An end-to-end entanglement is created as a result. An entanglement swapping operation consists of performing a joint BSM measurement with the qubits at the switch. }
\label{fig:quantum_switch_big_picture}
\end{figure}

One way to increase the number of qubits of a quantum computer is to connect multiple quantum processors \cite{MD16, guha2022cluster, qiao2022quantum} with a quantum switch. In brief, a quantum switch is analogous to a classic packet switch, but its task is to create end-to-end entanglements with the clients it is connected. Figure \ref{fig:quantum_switch_big_picture} shows an example of how a quantum switch operates. The switch first generates link-level entanglements (LLEs)\footnote{Also known as EPR pairs. A LLE or EPR pair consists of two entangled qubits \cite{YLS04}. One qubit at the switch and the other qubits at the client.} with the clients/processors (Figure \ref{fig:quantum_switch_big_picture}b), and then it uses these to create end-to-end entanglements (Figure \ref{fig:quantum_switch_big_picture}c \& d).\footnote{An end-to-end entanglement is created by performing a measurement (BSM or GHZ) on the qubits at the switch \cite{Nie02}. The process is also known as entanglement swapping when the requests are bipartite \cite{Pan98,JTT15}.} 
The end-to-end entanglements are used by the quantum applications  to, for example, teleport information qubits or carry out distributed quantum operations (via non-local CNOT gates  \cite{YLS04}). 

Quantum networking is in its infancy since single-hop communications are still challenging \cite{JTT15}. However, the building blocks of how quantum networks will operate already exist, prompting researchers to start designing the algorithms that will run the networks  when the hardware becomes available \cite{DT21, ADG+21, CCP21,LN22, panigrahy2022optimal}. 
Regarding quantum switches, previous work has studied their operation under a variety of settings  \cite{VGNT21,NVGT22, DRT21, VT22}. In brief, \cite{VGNT21} and \cite{NVGT22} study an idealized switch with bipartite and tripartite end-to-end entanglements requests when the request arrivals are symmetric and decoherence \cite{nielsen2002quantum} is negligible. The work in \cite{DRT21} studies a quantum switch with bipartite requests when there is no memory decoherence and LLE attempts succeed probabilistically. The contributions of  \cite{DRT21} are to characterize the switch capacity region and to propose on-demand policies that are throughput optimal. Similarly, the recent work in \cite{VT22} extends the setting in \cite{DRT21} to capture that LLEs expire (i.e., ``decohere'') after some time in practical systems.
In sum, previous work has focused \emph{primarily} on studying quantum switches for different decoherence models. However, none of them consider that quantum memory is a scarce resource that must be managed. In practice, quantum switches can only store a limited number of qubits (in analogy to quantum computers), constricting the LLEs that can exist at a given time and, therefore, the requests that the switch can serve.

In this paper, we study the problem of operating a quantum switch when it can store a limited number of qubits. In particular, the switch has fewer quantum memories than the number of clients it is connected, and so it has to decide how to allocate quantum memories to generate LLEs. 
 Studying this problem is important because memory is a scarce resource in practical quantum systems. 
To this end, this paper makes the following contributions:

\begin{itemize}
\item We present the first mathematical model of a quantum switch that has to operate with fewer quantum memories than the number of clients it is connected (Section \ref{sec:model}). Our model allows LLEs to decohere and end-to-end entanglement requests to be multipartite. 
\item We characterize the capacity region of the quantum switch with memory constraints (Section \ref{sec:capacity_region}), i.e., the set of end-to-end entanglement request arrival rates for which there exists a policy that can stabilize the switch.

\item We propose a memory allocation policy (MEW) that stabilizes the switch when (i) the LLEs last one time slot and (ii) the arrivals of end-to-end entanglement requests are in the interior of the capacity region  (Section \ref{sec:throughput_optimal_algorithm}). Finding a throughput optimal policy in this setting is challenging because the admissible scheduling decisions depend on the memory allocation. Such coupling is typically not allowed in classic networking problems (e.g., wireless) where the set of available actions can vary over time, in an i.i.d. manner \cite{GNT06} (see discussion after Theorem \ref{th:main_theorem}).

\item We present MEW2, a polynomial time variant of MEW tailored to the case where end-to-end entanglements are bipartite and LLE attempts are always successful. This case is important since multipartite requests can be divided into multiple bipartite requests (universality of two-qubit gates \cite{nielsen2002quantum}) and because, with sufficient entanglement distillation, LLEs attempts succeed almost surely \cite{pan2001entanglement}. 
\end{itemize}

Finally,  in Section \ref{sec:numerical_evaluation}, we evaluate MEW and MEW2 numerically depending on the requests arrivals characteristics and the time available to obtain a memory allocation.


\section{Quantum Switch Model \& Operation}
\label{sec:model}
\subsection{Switch model and operation overview}
\label{sec:model_operation}

\begin{figure}[t]
\centering
\includegraphics[width=0.6\columnwidth]{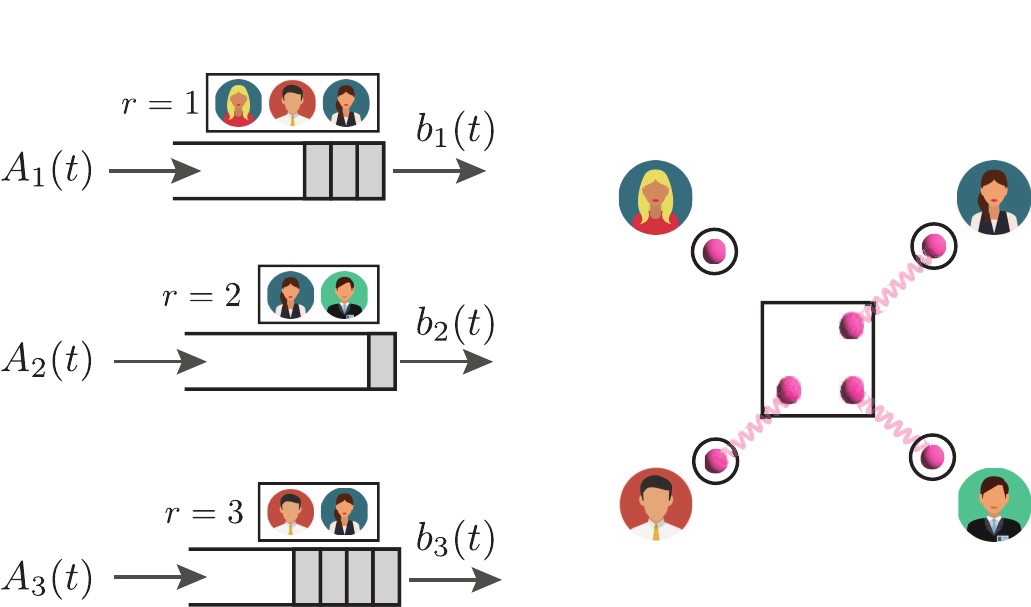}
\caption{Example of a quantum switch with three types of requests. The switch is connected to three users, which only allows it to serve requests of type 2 and 3.  }
\label{fig:switch_operation}
\end{figure}

We consider a quantum switch with $M$ quantum memories and $N$ clients that operates in slotted time. In each time slot $t = 1,2,3,\dots$, the switch receives a vector of requests
\begin{align*}
A(t) = (A_1(t), \dots, A_R(t)) ,
\end{align*}
where $A_r(t) \in \{0,1\}$ for all  $r \in \{1,\dots,R\}$. A request $A_r(t)$ involves connecting two or more clients (i.e., it is multipartite), and we use set
\begin{align}
\label{omega_def}
\Omega(r) \subseteq \{1, \dots, N\}
\end{align}
to denote the clients that participate in a request. For example, $\Omega(r) = \{1, 2\}$ if a request of type $r$ connects clients 1 and 2.

Upon arrival, the requests are stored in separate queues $Q(t) = (Q_1(t),\dots,Q_R(t))$ to await service. The queues evolve as follows:
\begin{align}
\label{queue_evolution}
    Q(t+1) = [Q(t) - b(t)]^{+} + A(t),
\end{align}
where $[\cdot]^+ := \max\{0,\cdot\}$ and 
\[
b(t)  = (b_1(t),\dots,b_R(t))
\] 
 indicates the requests served in time slot $t$. In particular, $b_r(t) =1$ if a request $r \in \{1,\dots,R\}$ is served, and  $b_r(t) =0$ otherwise. 

The switch's task is to serve as many requests as possible subject to operational constraints. In particular, the switch can only serve a request if all the clients that participate in it have an active LLE. Figure~\ref{fig:switch_operation} shows an example of a switch with four clients and three types of requests $r \in \{1,2,3\}$. Observe from the figure that the switch can serve requests of type 2 and 3, but not of type 1 because one of the clients is not connected with the switch. 

In the next section, we describe how the switch allocates quantum memories to clients and how that affects the switch connectivity and the set of admissible service vectors.\footnote{i.e., the requests that can be served given a switch connectivity.}
\subsection{Switch operation and decision variables}
\label{sec:quantum_operation_decription}

\begin{figure}[t]
\centering
\includegraphics[width=0.6\columnwidth]{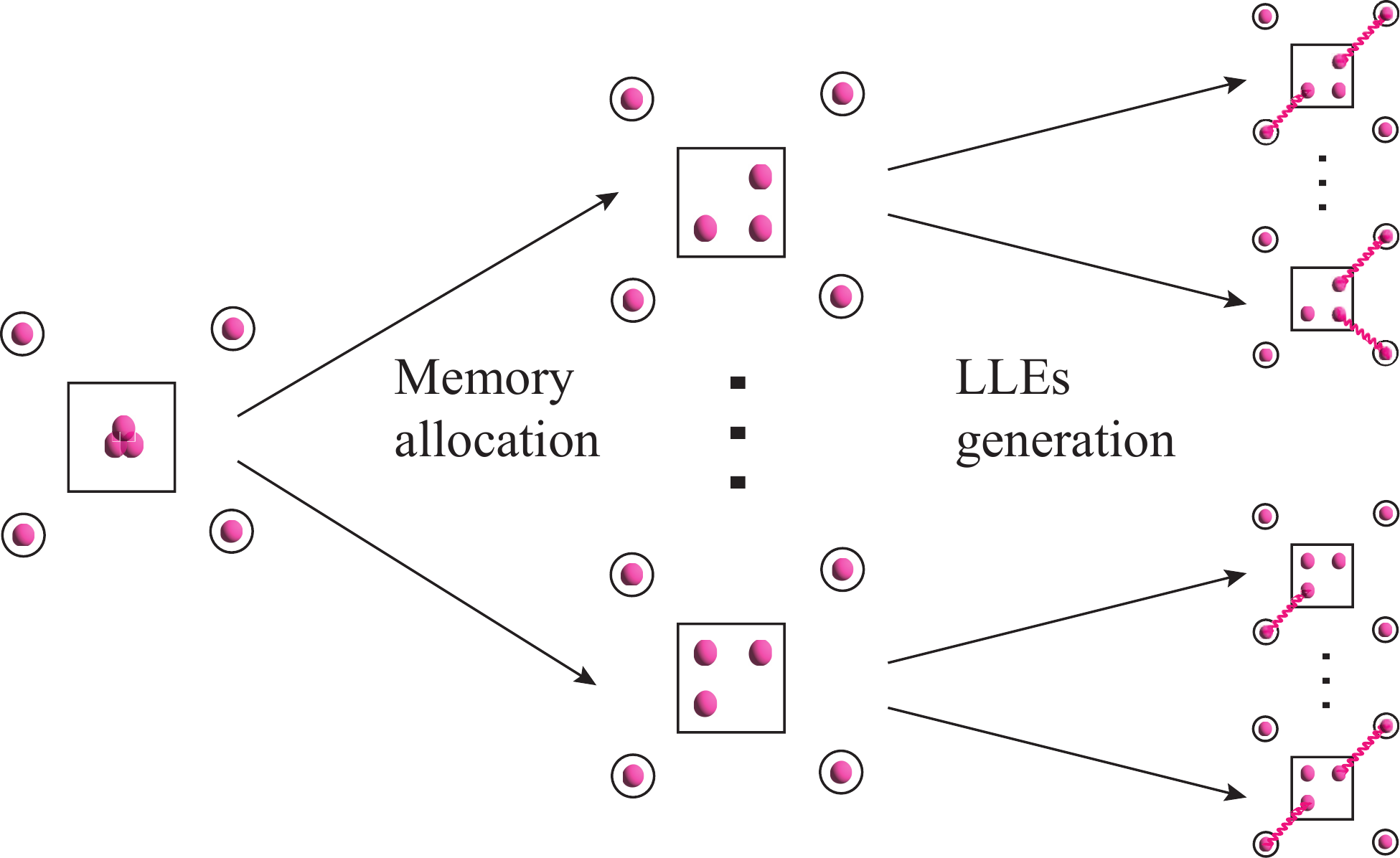}
\caption{ Illustrating how the quantum memory allocation results in different possible connectivities in a quantum switch with $M =3$ and $N=4$. Observe from the figure that different memory allocations can result in the same switch connectivity.} 
\label{fig:switch_possibilities}
\end{figure}

In each time slot, the quantum switch performs three types of actions. It
(i) allocates quantum memories to clients;
(ii) generates LLEs; and 
(iii) serves multipartite requests by using the LLEs. Importantly, a LLE can only be used to serve one request as this is consumed to generate an end-to-end entanglement \cite{NVGT22}.
Next, we describe the control variables that the switch can select in each time slot. 

\subsubsection{Quantum memory allocation}
When $M < N$, the switch has to decide how to allocate memories to clients. We use $m_n(t)$ to denote whether the switch assigns a quantum memory to a node $n \in \{1,\dots,N\}$ in time slot $t$, and collect these in vector
\[
m(t) = (m_1(t), \dots, m_N(t)).
\]
The set of eligible memory allocations $\mathcal{M}$ is given by
\begin{align*}
\mathcal{M} & = \Bigg\{ (m_1, \dots, m_N): m_n \in \{ 0,1\}\   \forall n \in \{1,\dots,N\} \\
& \qquad \qquad \qquad \quad \qquad\text{with} \ \  \sum_{n = 1}^{N} m_n \le M \Bigg\}.
\end{align*}
That is, $\mathcal M$ contains the binary vectors whose components' sum is smaller than or equal to the number of memories available (i.e., $M$). 

\subsubsection{LLEs generation and switch connectivity} 
After the memory allocation, the switch has to generate LLEs with the clients that are connected to a memory. The switch attempts to create LLEs by sending entangled qubits (e.g., photons) over a fiber-optical channel, but only a fraction of the LLE attempts are successful due to interference \cite{YFL20}. Also,  LLEs last for a limited amount of  time due to a phenomenon known as decoherence \cite{Bac20}.

We model the switch connectivity in a time slot as follows. Let $p_n \in [0, 1]$, $n \in \{1,\dots,N\}$ be the probability that a LLE attempt succeeds. Vector 
\[
k(t) = (k_1(t), \dots, k_N(t))
\]
with 
\begin{align}
k_n(t) = 
\begin{cases}
0, & m_n(t) = 0, \\
0, & m_n(t) = 1 \  \text{ w.p.} \ \  1-p_n\\
1, & m_n(t) = 1 \ \text{ w.p.} \ \ p_n
\end{cases}
\end{align}
denotes the collection of successful LLEs in a time slot, i.e., the switch's connectivity. We use set 
\begin{align}
\mathcal{K}(m(t)) \subseteq \{0,1\}^N
\label{eq:cases_k}
\end{align}
to capture all the  possible switch connectivities for a given memory allocation $m(t) \in \mathcal M$. Note that a memory allocation has a total of $| \mathcal{K}(m)| = 2^M$ possible switch connectivites if all the memories are used.\footnote{Otherwise, the switch has $2^{\sum_{r=1}^R m_r(t)}$ possible connectivities.}   
Figure \ref{fig:switch_possibilities} shows how the switch connectivity depends on different memory allocations and the successful LLEs. Also, observe from the figure that different memory allocations can result in the same connectivity due to some LLE attempts failing. 

The duration of a switch connectivity depends on how long LLEs last. In this paper, we make the following assumption: 
\begin{assumption}
LLEs last for one time slot.
\label{as:time-slot-decoherence}
\end{assumption}

This assumption is standard (e.g., \cite{VT22}), and it allows us to align the duration of a LLE with the frequency in which the switch allocates quantum memories and serves requests. 

\subsubsection{End-to-end entanglement requests service} The switch connectivity in a time slot affects the set of available service vectors. Let $k(t) \in \mathcal{K}(m(t))$ with $m(t) \in \mathcal M(t)$ be the switch connectivity at time slot $t$. The set of admissible service vectors is given by:\footnote{Although the set $\mathcal{B}(m(t), k(t))$ depends only on the network connectivity, $k(t)$, we parameterize it with $m(t)$ as well to emphasize that the service vectors are picked after the memory allocation. }

\small
\begin{align*}
\mathcal{B}(m(t), k(t)) & =  \Bigg\{ b_r \in \{0,1\}, r \in \{1,\dots,R\} : 
\\
& \qquad \text{there exists a matrix } S \in \{0,1\}^{R\times N} \\
& \qquad \text{s.t.\ } s_{rn} = 1 \text{ for all } n \in \Omega(r) \text{ iff } b_r=1, \\
&  \qquad \text{and } \sum_{r=1}^{R} s_{rn} \le k_n (t) \ \forall n\in\{1,\dots,N\}
\Bigg\}.
 \end{align*}
 \normalsize
That is, $\mathcal B(m(t),k(t))$ contains a collection of binary vectors, where the  $r$'th entry of a vector is equal to one if and only if (i) all the clients involved in a request of type $r$ have an active LLE with the switch, and (ii) a LLE is only used to serve one request.

\section{Capacity Region and \\
 Throughput Optimal Policy}

In this section, we present the main contributions of the paper: the characterization of the capacity region of the quantum switch (Section \ref{sec:capacity_region}),  and a memory allocation policy that is  throughput optimal (Section \ref{sec:throughput_optimal_algorithm}). In Section \ref{sec:novelty_and_complexity}, we discuss the scalability of the proposed policy.

\subsection{Capacity region}
\label{sec:capacity_region}

Before designing an algorithm, we need to characterize the set of arrival rates that the switch can support. To start, let 
\begin{align}
\lambda := \lim_{T\to \infty} \frac{1}{T} \sum_{t=1}^T A(t)
\end{align}
be the long-term arrival rate of requests at the quantum switch. We say an arrival vector $\lambda$ is admissible (or, it can be supported) if there exists a policy $\pi$ that can generate a sequence of service rate vectors $\{ b^\pi(t) \}_{t = 1}^\infty$ such that
\begin{align}
\lambda_r \le f^\pi_r : = \lim_{T \rightarrow \infty} \frac{1}{T} \sum_{t = 1}^T b_r^\pi(t)
 && \forall r \in \{1,\dots,R\}. \label{eq:basic_inout_ineq}
\end{align}
That is, for a given vector $\lambda$, the switch must be able to generate a long-term service vector $f^\pi$ that is equal to or larger than $\lambda$ component-wise. Hence, by characterizing all the service vectors $f^\pi$ that the switch can generate, we know the arrival vectors that the switch can support (i.e., the switch capacity region).

To define the switch capacity region, we decouple the decision variables from the time slot index $t$ and express them as the fraction of time they can occur. 
In short, let $\theta_m$ denote the fraction of time a memory allocation $m \in \mathcal{M}$ is used, and $\mathbb{P}(k; m)$ the probability that a switch connectivity $k \in \mathcal K(m)$ occurs for a given a memory allocation $m \in \mathcal M$.   
Similarly, let $\delta_b^{k,m}$ be  the fraction of time that each service vector $b \in \mathcal{B}(m, k)$ is used for a given switch connectivity and memory allocation. We have the following proposition.

\begin{proposition}[Quantum switch capacity region]
\label{th:capacity}
\label{capacity_region}
The capacity region of the quantum switch is:
\small
\begin{align}
\Lambda :=   
\Bigg\{  
& f^\pi: f^\pi = \sum_{m \in \mathcal{M}} \theta_m \sum_{k \in \mathcal{K}(m)}  \mathbb{P}(k; m) \sum_{b \in \mathcal{B}(m,k)} \delta^{m,k}_{b} b, \notag \\ 
& \quad \sum_{m \in \mathcal{M}}\theta_m = 1, \sum_{b \in \mathcal{B}(m, k)} \delta^{m,k}_{b} = 1, \notag \\ 
& \quad \theta_m \ge 0,  \ \delta^{m,k}_{b} \ge 0, \notag \\ 
& \quad \textup{for all } b \in \mathcal{B}(m,k),  \ k \in \mathcal{K}(m), \ m\in \mathcal{M} \Bigg\}.
\end{align}
\normalsize
%
%
\end{proposition}
\begin{IEEEproof}[Proof sketch] The full proof is omitted due to space constraints. However, it follows the same methodology as in \cite{GNT06, TE92}: writing the fraction of time that the service vectors can be generated---depending on the memory allocations and switch connectivities in our case. 
\end{IEEEproof}

Note that if $\lambda \in \Lambda$ (i.e., the long-term average of requests arrivals is in the capacity region), then there exists a vector $f^\pi$ that satisfies \eqref{eq:basic_inout_ineq}. Having $\lambda \in \Lambda$ is usually known as the \emph{necessary} condition for having stable queues \cite{GNT06}.

\subsection{MEW: A max-weight algorithm for allocating quantum memory and serving requests}
\label{sec:throughput_optimal_algorithm}

\begin{algorithm}[t!]
\small
\begin{algorithmic}[1]
\STATE \textbf{Set:} $t = 0$
\WHILE{switch is operating}
\STATE $t \leftarrow t + 1$ 
\STATE \textbf{(S1) Quantum memory allocation:} Select the memory allocation
\begin{align}
\label{eq:memoryallocation}
m(t) \in \underset{{m \in \mathcal{M}}}{\arg\max} \sum_{r = 1}^R Q_r(t) \mu_r(m,Q(t)),
\end{align}
where 
\begin{align}
\mu (m, Q(t)) & : = \sum_{k \in \mathcal{K}(m)} \mathbb P(k) w(k, Q(t))  \label{eq:expected_service_state} \\
w(k, Q(t)) &  \in \underset{{u \in \mathcal B(m, k) } }{\arg \max} \sum_{r = 1}^R Q_r(t) u_r.
\label{eq:mw-update_bsm}
\end{align} 
\STATE \textbf{(S2) LLEs generation:} The switch attempts to create LLEs with the clients that have a memory connected. The successful LLEs determine the switch connectivity $k(t)$ and the action set $\mathcal B(m(t),k(t))$. 
\STATE \textbf{(S3) Requests service:} Select a service vector with the update
\[
b(t) \in \underset{{u \in \mathcal B(m(t),k(t))}}{\arg \max} \sum_{r = 1}^R Q_r(t) u_r 
\]

\STATE \textbf{(S4) Queue update:} Serve end-to-end entanglement requests and update the queues with arrivals $A(t)$:
\begin{align}
Q(t+1) = [Q(t) - b(t) ]^+ + A(t) 
\label{eq:queue_update_alg}
\end{align}
\ENDWHILE
\end{algorithmic}
\caption{(MEW)}
\label{al:max-weight}
\end{algorithm}

We present \textit{Maximum Expected Weight (MEW)}, an algorithm that stabilizes the queues  when the long-term arrival of requests is in the \emph{interior} of the capacity region.
MEW (Algorithm \ref{al:max-weight}) consists of three steps. The first step \textbf{(S1)} allocates the quantum memories to clients using \eqref{eq:memoryallocation}, which consists of maximizing the sum of the expected service in each queue (i.e., $\mu_r$) multiplied by the queue occupancies (i.e., $Q_r$). This update can be regarded as an ``expected'' max-weight maximization, where the updates in \eqref{eq:expected_service_state} and \eqref{eq:mw-update_bsm} are intermediate steps to compute the expected rate vectors $\mu(m,Q(t)) = (\mu_1(m,Q(t)),\dots,\mu_R(m,Q(t)))$ used in   \eqref{eq:memoryallocation}. 
The second step \textbf{(S2)}  generates the LLEs with the clients that have a memory connected. Only some LLE attempts succeed due to interference, which affects the network connectivity and the set of admissible requests service vectors, i.e., set $\mathcal B(m(t),k(t))$. 
The third step \textbf{(S3)} consists of finding the service vector $b(t)$  that maximizes the  dot product with the vector of queues $Q(t)$. Once all the decision variables have been made, the queues are updated as indicated in \eqref{eq:queue_update_alg}. 
We have the following theorem.

\begin{theorem}
\label{th:main_theorem}
Consider the quantum switch model in Section \ref{sec:model}, and suppose that the long-term arrival rate of requests $\lambda$ is in the interior of the capacity region $\Lambda$. That is, there exists a vector $\hat b \in \Lambda$ such that %
\begin{align*}
\lambda_r + \epsilon \le \hat b_r && \forall r \in \{1,\dots,R\}
\end{align*}
for some $\epsilon > 0$. 
Then, MEW (Algorithm \ref{al:max-weight}) ensures that
\begin{align*}
\lim_{T \to \infty} \frac{1}{T} \sum_{t=1}^T \sum_{r=1}^R  \mathbb E[Q_r(t)]
& \le  \frac{N^2}{\epsilon},
\end{align*}
i.e., the queues are strongly stable. 
\end{theorem}

\begin{IEEEproof}
See Section \ref{sec:appendix_proof_thm1}.
\end{IEEEproof}

Strong stability implies that all the requests that arrive are eventually served (i.e.,  \eqref{eq:basic_inout_ineq} is satisfied), but also that the queues are bounded \cite{neely2010stability}.
The result in Theorem \ref{th:main_theorem} is based on max-weight techniques widely employed in network scheduling problems \cite{TE92, GNT06}, and the novelty of our contribution resides in the fact that the switch connectivity is random and depends on how we assign quantum memories to links/clients. The latter is different from wireless network models with time-varying connectivity since the allocation of quantum memories affects the switch's connections and, therefore, the set of admissible service vectors. Such coupling is typically not allowed in max-weight or backpressure approaches where the set of available actions can vary over time; however, usually in an i.i.d. manner \cite{Tas97}. In our problem, the action sets $\{\mathcal B(m(t),k(t))\}_{t=1}^\infty$ are not i.i.d.\ because they depend on the memory allocation decisions $\{m(t) \in \mathcal M \}_{t=1}^\infty$. Our approach to tackle this problem is to exploit the linearity of  \eqref{eq:memoryallocation}, \eqref{eq:expected_service_state}, and \eqref{eq:mw-update_bsm}, and evaluate all the possible scheduling decisions for every connectivity. However,  enumerating all the cases can be computationally expensive sometimes, as we discuss next.

\subsection{MEW scalability }
\label{sec:novelty_and_complexity}

The step with higher computational cost is the allocation of quantum memories  \textbf{(S1)}, which involves computing \eqref{eq:memoryallocation}, \eqref{eq:expected_service_state}, and \eqref{eq:mw-update_bsm}.
In brief, the maximization in \eqref{eq:memoryallocation} is over the set of all possible memory allocations, which has cardinality $|\mathcal{M}| = {N \choose M}$.\footnote{Assuming we allocate all the memories to clients.} Furthermore, we need to compute \eqref{eq:expected_service_state} and \eqref{eq:mw-update_bsm} for every memory allocation $m \in \mathcal M$ and network connectivity $k \in \mathcal{K}(m)$ respectively.
We could compute \eqref{eq:mw-update_bsm} only once per switch connectivity since a switch connectivity can be obtained by different quantum memory allocations (see example in Figure \ref{fig:switch_possibilities}). However, the number of possible switch connectivities increases exponentially with the number of memories since $|\mathcal{K}(m)| = 2^M$. In addition, the update in  \eqref{eq:mw-update_bsm} requires finding a maximum-weighted matching in a complete hypergraph,\footnote{A hypergraph is a generalization of a graph in which an edge can connect any number of vertices.} which is known to be an NP-hard problem \cite{HT16}.

In sum, MEW does not scale well with $N$ and $M$ since it solves, in the worst case, an exponential number of NP-hard problems for every memory allocation. However, MEW can be used effectively when $N$, $M$ and $R$ are not very large. In the next section, we focus on a special case where we can derive a variant of MEW (MEW2) that has polynomial complexity.


\section{MEW2: Efficient Scheduling for Bipartite Requests and Successful LLEs}
\label{sec:variant}

\begin{figure}[t]
\centering
\includegraphics[width=0.7\columnwidth]{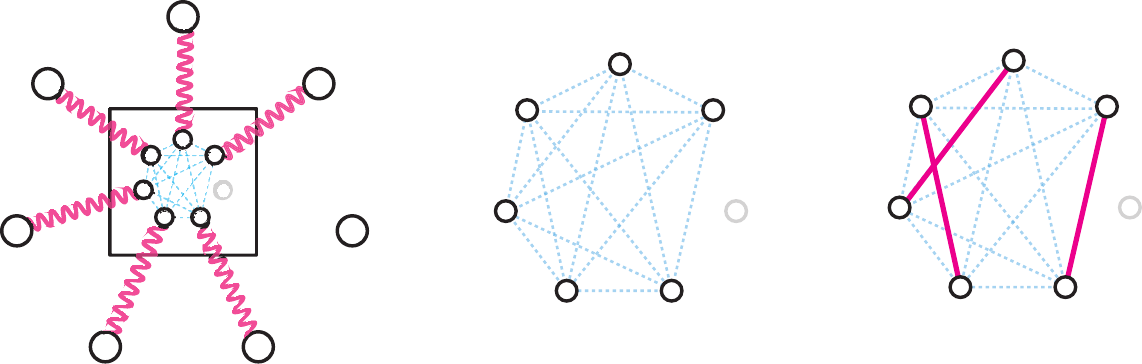}
\caption{Quantum switch with $N= 7$ clients and $M= 6$ memories. When requests for end-to-end entanglements involve only two clients, the update in \eqref{eq:mw-update_bsm} reduces to finding a maximum weighted matching in a complete graph.}
\label{fig:bimatching_ex}
\end{figure}

\begin{figure}[t!]
\centering
\includegraphics[width=0.85\columnwidth]{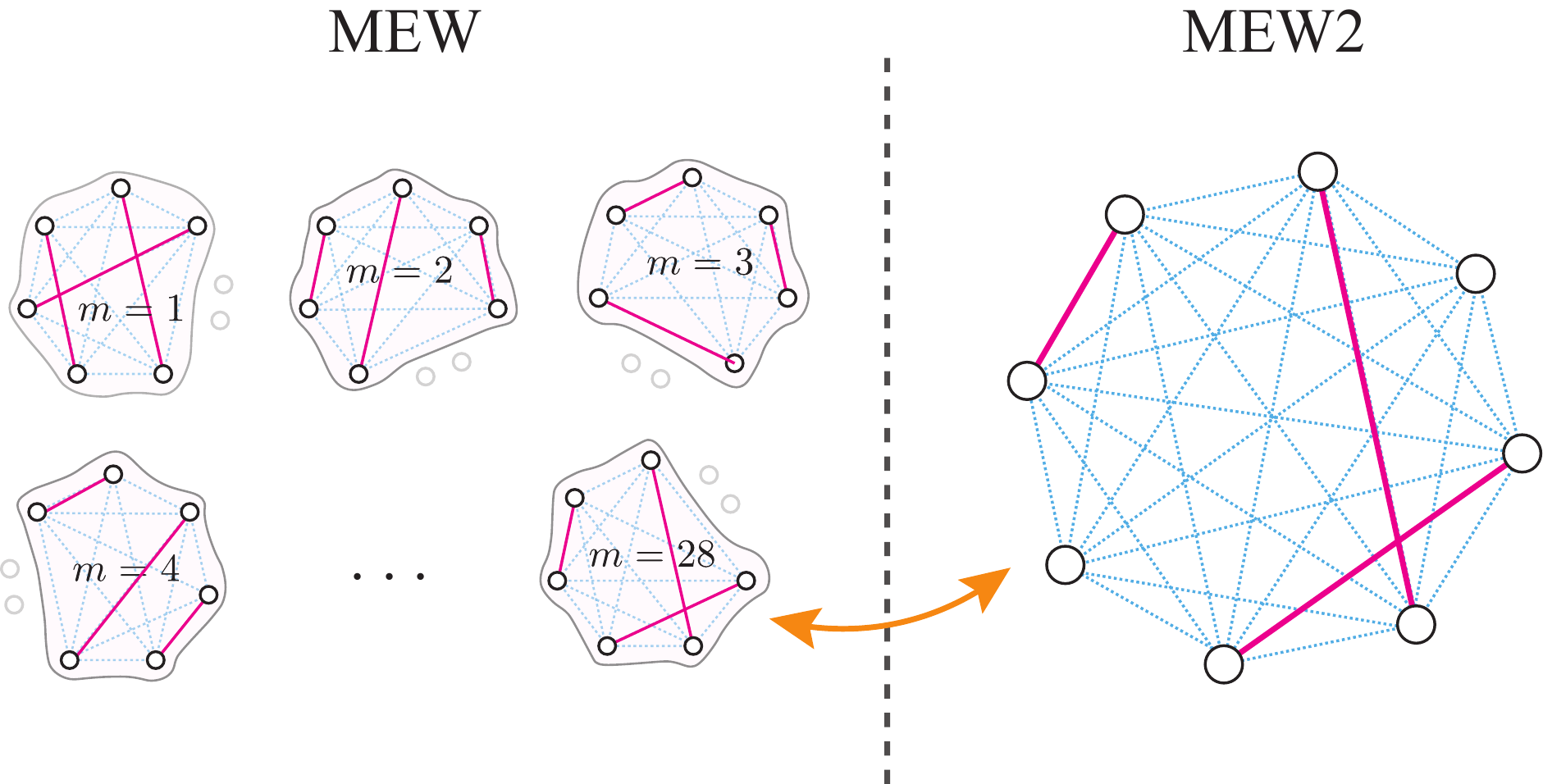}
\caption{Schematic illustration of how MEW compares to MEW2 for a switch with $N =8$ clients and $M=6$ quantum memories. MEW has to find a maximum weighted matching in each of the ${8 \choose 6} = 28$ different complete graphs (with 6 nodes each). 
MEW2 selects a special type of matching with at most $M/2$ edges in the $8$-complete graph.}
\label{fig:alg1vsalg2}
\end{figure}

In this section, we study the case where:
\begin{itemize}
\item (i) LLE attempts are always successful (i.e., $p_n = 1$, $ \forall n \in \{1,\dots,N\}$), and 
\item (ii) Requests involve only two clients.\footnote{Without loss of generality we assume that $M$ is even. With bipartite requests, having an odd number of memories means that there will be an unused memory.}
\end{itemize} 
This case is important because of two reasons. First, we can extend the duration of a time slot to perform entanglement distillation and increase the probability that every LLE succeed.\footnote{Note that there is a trade-off between reducing the duration of a time slot (thus increasing the number of service requests per unit of time), and increase the probability that LLEs succeed.} Second, every multipartite request between clients can be divided into (multiple) bipartite ones. That is because two-qubit gates are universal, i.e., every quantum program can be implemented with two-qubit gates \cite{nielsen2002quantum}.

This case allows us to derive a variant of MEW (MEW2) that has lower computational cost. Specifically, we can allocate memories and select which requests to serve by finding a special type of matching with at most $M/2$ edges in an $N$-complete graph (see Algorithm \ref{al:low-complexity-algorithm}).\footnote{An $N$-complete graph is a graph with $N$ nodes, in which each pair of graph vertices is connected with an edge. The edges' weights are the queue backlogs.}

\subsection{Motivation: Complexity of MEW}

When end-to-end entanglement requests involve only two clients,  \eqref{eq:mw-update_bsm} corresponds to finding a maximum weighted matching in the complete graph of the clients with an active LLE (see Figure \ref{fig:bimatching_ex}). Finding such matching has polynomial time complexity (see, for example, the survey in \cite{schrijver2003combinatorial}). 
The assumption that LLEs are always successful is useful to reduce the number of times we need to call  \eqref{eq:mw-update_bsm}. In particular, we have that a memory allocation is associated with a \emph{single} switch connectivity. Hence, $|\mathcal K(m(t)) |= 1$ for all $m(t) \in \mathcal M$ and 
\[
\mu(m(t),Q(t)) = w(k(t),Q(t))
\]
since $m(t) = k(t)$. In sum, we can solve \eqref{eq:mw-update_bsm} in polynomial time and only once for every admissible memory allocation. Yet, that can still be too much in some cases.  For example, if $N = 16$ and $M  = 8$, we need to find a maximum weighted matching of ${16 \choose 8} = 12,870$ different graphs to make a single memory allocation decision. 

\subsection{Maximum Expected Weight 2 (MEW2)}

We propose \textit{Maximum Expected Weight 2 (MEW2)}, a policy that selects a memory allocation by obtaining 
a special type of matching in the $N$-complete graph (Algorithm \ref{al:low-complexity-algorithm}). 
The intuition behind MEW2 is shown in Figure \ref{fig:alg1vsalg2} for a switch with $N=8$ clients and $M= 6$ memories. Recall that \textbf{(S1)} in  MEW computes a maximum weighted matching in $N \choose M$ different $M$-complete graphs, and picks one that has maximum weight.  Observe from the figure that such matching is also a (non-maximal) matching in the $N$-complete graph. Thus, we can replace step \textbf{(S1)} in MEW by \emph{directly} computing a matching with maximum weight among the matchings that have at most $M/2$ edges. We have the following corollary of Theorem \ref{th:main_theorem}.

\begin{algorithm}[t]
\small
\begin{algorithmic}[1]

\STATE \textbf{Set:} $t = 0$
\WHILE{switch is operating}
\STATE $t \leftarrow t + 1$  
\STATE \textbf{(S1b) Quantum memory allocation:} Select a matching with at most $M/2$ edges in the $N$-complete graph with maximum possible weight:
\begin{align}
\label{low_b}
l(t) \in \underset{{ u \in O }}{\arg\max}  \ \sum_{r=1}^R Q_r (t) \ u_r,
\end{align}
where $O := \{ u \in \mathcal{P}_N: \sum_{r=1}^R u_r \le M/2 \}$ and $\mathcal{P}_N$ the set of matchings in the $N$-complete graph.

Assign a memory to every client/node that is connected to an edge in $l(t)$, i.e., 
\begin{align}
\label{low_m}
m(t) \in \{m \in \mathcal{M}: l(t) \in \mathcal{P} ( \mathcal{C}(m) )\},
\end{align}
where $\mathcal{C}(m)$ is the complete graph of the clients $n \in \{1,\dots,N\}$ with $m_n = 1$.
\STATE \textbf{(S2b) LLE generation:} Generate LLEs with the clients that have a memory connected. 
\STATE \textbf{(S3b) Requests service:} Select
\begin{align*}
b(t) = l(t)  
\end{align*}
\STATE \textbf{(S4b) Queue update:} 
\begin{align}
Q(t+1) = [Q(t) - b(t) ]^+ + A(t) 
\end{align}
\ENDWHILE
\caption{(MEW2)}
\label{al:low-complexity-algorithm}
\end{algorithmic}
\end{algorithm}

\begin{corollary}[Theorem \ref{th:main_theorem}]
\label{cor:1}
Consider the setup of Theorem \ref{th:main_theorem} where the LLE attempts are always successful (i.e., $p_n = 1, \forall n \in \{1,\dots,N\}$). Also, suppose that requests involve connecting two clients and that  $M$ is even. Then, MEW2 ensures that the queues are strongly stable. 
\end{corollary}
\begin{IEEEproof}
See Appendix (Section \ref{sec:proof_coro1}). 
\end{IEEEproof}

Finding the matching described in \eqref{low_b}, which characterizes the complexity of MEW2, can be done in polynomial time. 
%
%
In particular, we can find such matching by augmenting the $N$-complete graph and then computing a maximum weighted matching. Specifically, the augmented graph has $n-M$ virtual nodes connected to the others with edges that have \textit{infinite} weight. The solution to \eqref{low_b} corresponds to a maximum weighted matching of the augmented graph, which can be found in polynomial time \cite{edmonds1965paths, schrijver2003combinatorial}.

\section{Numerical Evaluation}
\label{sec:numerical_evaluation}

In this section, we perform three different simulations to evaluate the performance of MEW and MEW2. Our goal is to illustrate the algorithms' behavior in different scenarios (e.g., request load, LLEs generation) and to study MEW when the update in \eqref{eq:memoryallocation} is carried out approximately. In particular, when MEW uses only $l$ memory allocations out of all the $N \choose M$ possibilities (Sections \ref{sec:exp2} and \ref{sec:exp3}). We refer to such algorithm as $l$-Approximate MEW.


\subsection{Simulation 1: Performance of MEW under different arrival rates}
\label{sec:exp1}
\begin{figure}[t]
\centering
\includegraphics[width=0.7\columnwidth]{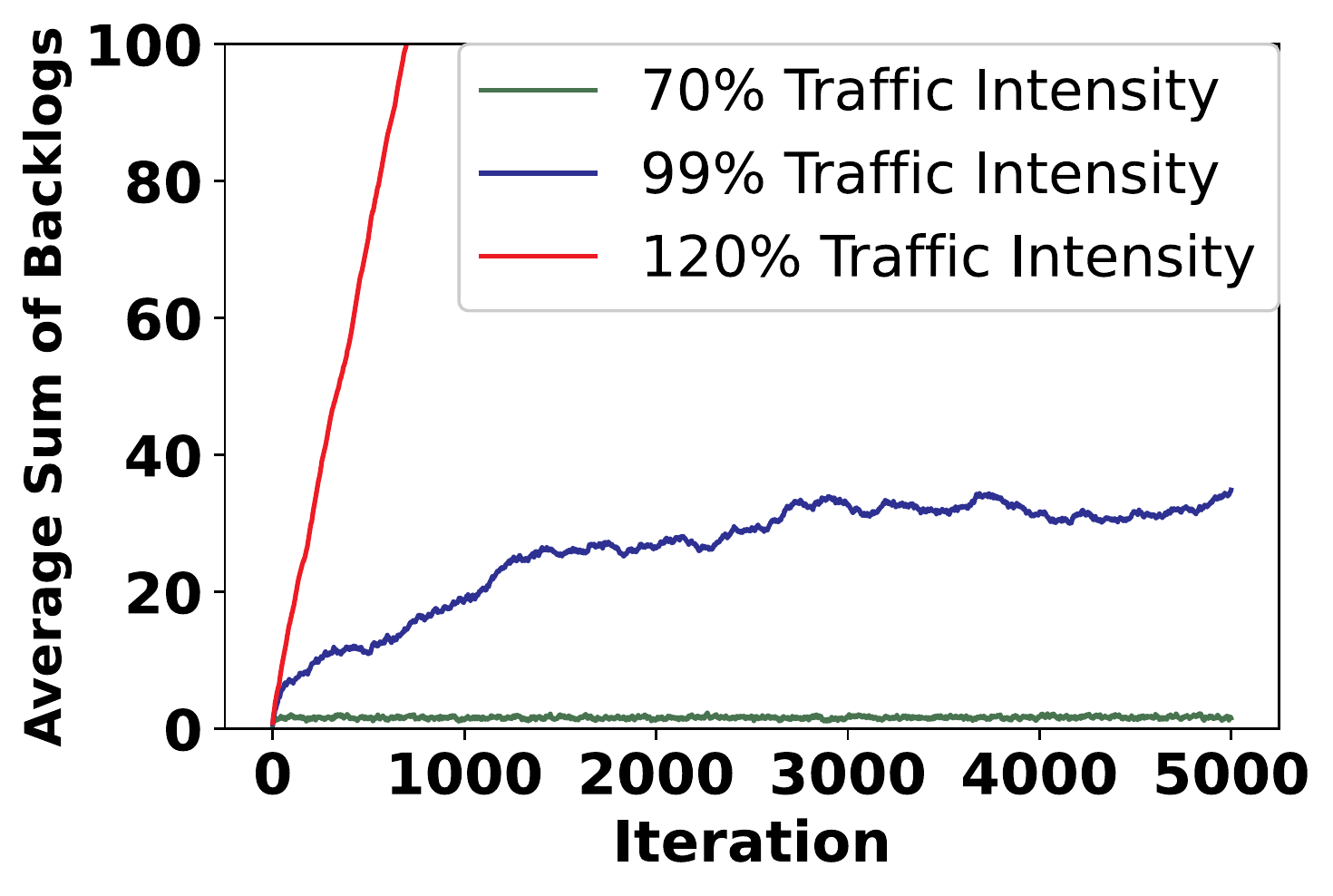}
\caption{Illustrating the simulation in Section \ref{sec:exp1}: The evolution of MEW for a quantum switch with $6$ clients and $3$ memories for different arrival rates. LLE attempts succeed with probability $0.9$. Each line is the average of $10$ different realizations.}
\label{fig:exp1}
\end{figure}

This simulation evaluates the performance of MEW when the requests arrive with three different intensities: $70 \%$, $99 \%$, and  $120 \%$ of the total load that the switch can support.\footnote{An intensity of $100\%$ is at the boundary of the capacity region.} For the simulation, we set $N = 6$, $M = 3$, and $R = 8$, where all the requests involve connecting three clients (i.e., the requests are tripartite). The probability of the LLE attempts being successful is fixed to $p_n = 0.9$ for all clients and loads.

We run MEW for the three different loads and show the evolution of the queue occupancies over time in Figure \ref{fig:exp1}. 
Observe from the figure that when the arrival rates are in the interior of the capacity region ($70\%$ and $99\%$), the backlogs remain bounded. However, note that the ``saturation'' points are different, which is in line with the queue stability bound in Theorem \ref{th:main_theorem}. Higher intensity (i.e., smaller $\epsilon$ in Theorem \ref{th:main_theorem}) implies larger backlogs. 
Finally, observe from the figure that when the request arrival intensity is equal to $120\%$ (outside of the capacity region), the queues are not stable since their occupancy increases linearly.

\textbf{Conclusions:} MEW  stabilizes the queues when the arrivals are in the interior of the capacity region. The average queue occupancies depend on how close the long-term arrival rates are to the boundary of the capacity region (Theorem \ref{th:main_theorem}).

\begin{figure}[t]
\label{exp2}
\centering
\begin{tabular}{c}
\includegraphics[width=0.7\columnwidth]{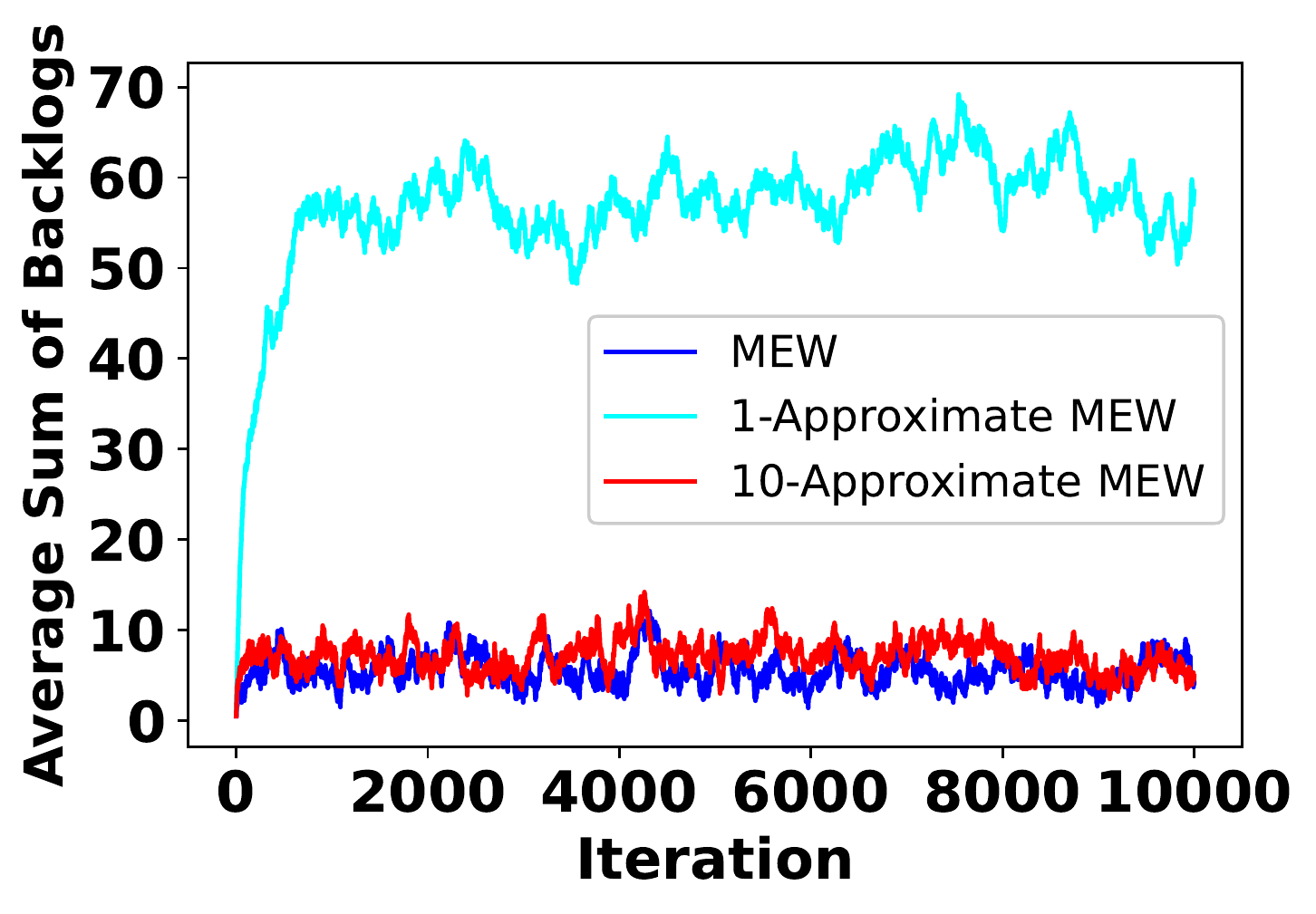}
\\
(a) 70\% Traffic intensity \\
\includegraphics[width=0.7\columnwidth]{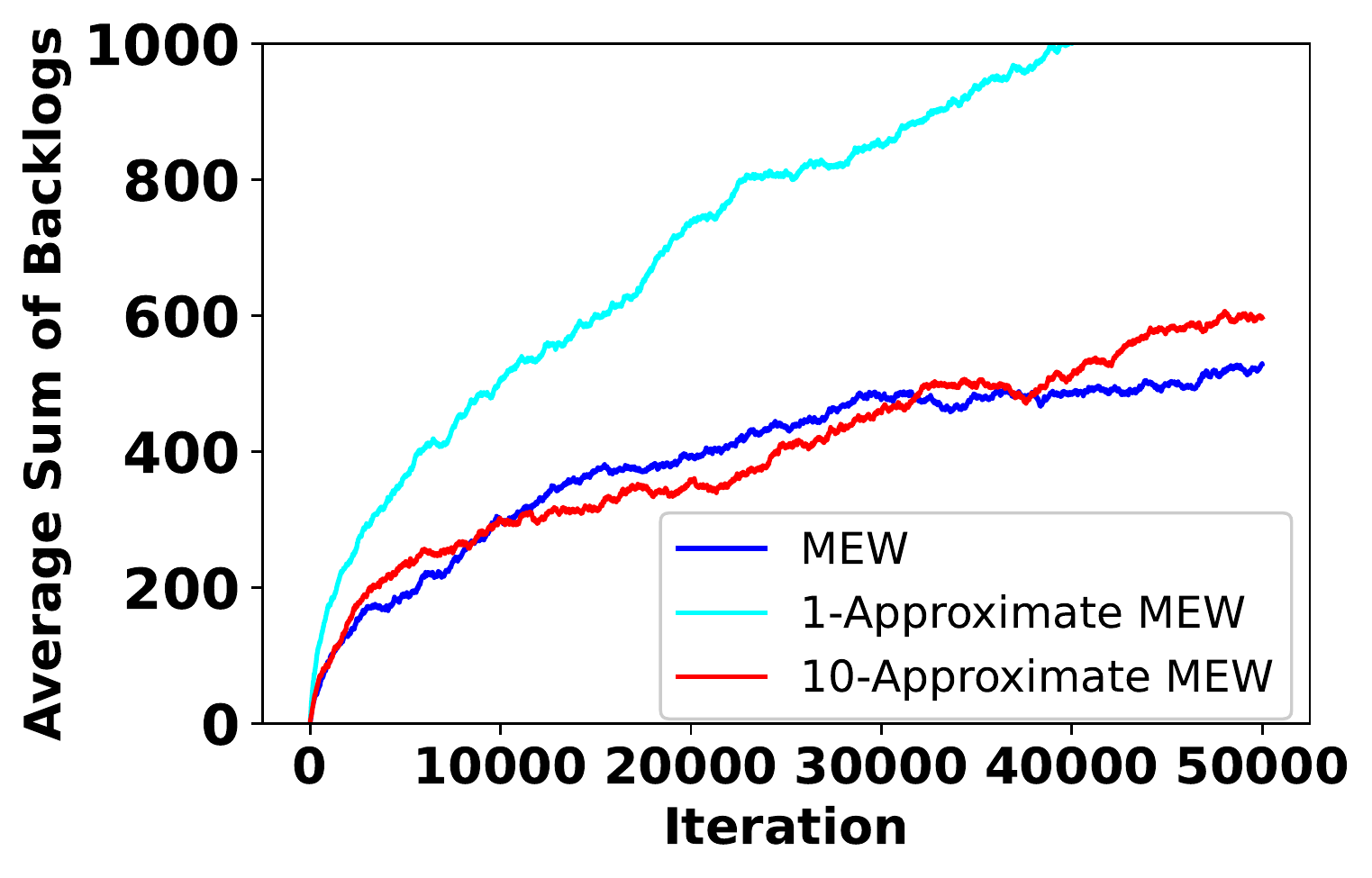} \\
(b) 99\% Traffic intensity
\end{tabular}
\caption{Illustrating the simulation in Section \ref{sec:exp2}: The evolution of MEW and $ \{1,10\}$-Approximate MEW for a quantum switch with $N = 6$, $M = 3$ and $p_n = 0.9$ for every client $n$. Each line is the average of $10$ different realizations.  }
\label{fig:exp2}
\end{figure}

\subsection{Simulation 2: MEW with memory allocation decision deadlines}
\label{sec:exp2}

Recall from Section \ref{sec:novelty_and_complexity} that MEW needs to solve \eqref{eq:mw-update_bsm} multiple times for every memory allocation. However, we may not be able to evaluate all possible memory allocations since  in practice we need to select one within a time deadline. To capture that, we reduce the search space of the problem in  \eqref{eq:memoryallocation}.

As in Section \ref{sec:exp1}, we consider $N = 6$ clients, $M = 3$ memories, and probabilities for successful LLE attempts equal to $p_n = 0.9$ for all clients. However, we now consider all types of bipartite and tripartite requests (i.e., $R = {6\choose2} + {6\choose3} = 35$), hence the time needed to compute \eqref{eq:mw-update_bsm} increases.

We run MEW and the  $\{1,10\}$-Approximate variant for different arrival rate intensities ($70 \%$, $99 \%$, and $120 \%$) and show the results in Figure \ref{fig:exp2}. Observe from the figure that MEW stabilizes the queues when the arrivals are in the interior of the capacity region. However, for the $l$-approximation, the stability depends on the value of $l$ and the traffic intensity. Specifically, the 1-Approximate MEW stabilizes the system when the traffic intensity is $70\%$ (Figure \ref{fig:exp2}a), but not when the intensity is equal to  $99\%$ (Figure \ref{fig:exp2}b). In contrast, the 10-Approximation keeps the queues bounded similar to MEW.

\textbf{Conclusions:} The $l$-Approximate MEW can stabilize the queues when the value of $l$ is large enough. How large $l$ should be is related to how close the arrival rates are to the boundary of the capacity region. As future work, it is interesting to investigate how the capacity region scales when the memory allocation is obtained approximately (e.g., as a function of the parameter $l$).

\subsection{Simulation 3: Performance of MEW2 }
\label{sec:exp3}

\begin{figure}[t]
\centering
\includegraphics[width=0.7\columnwidth]{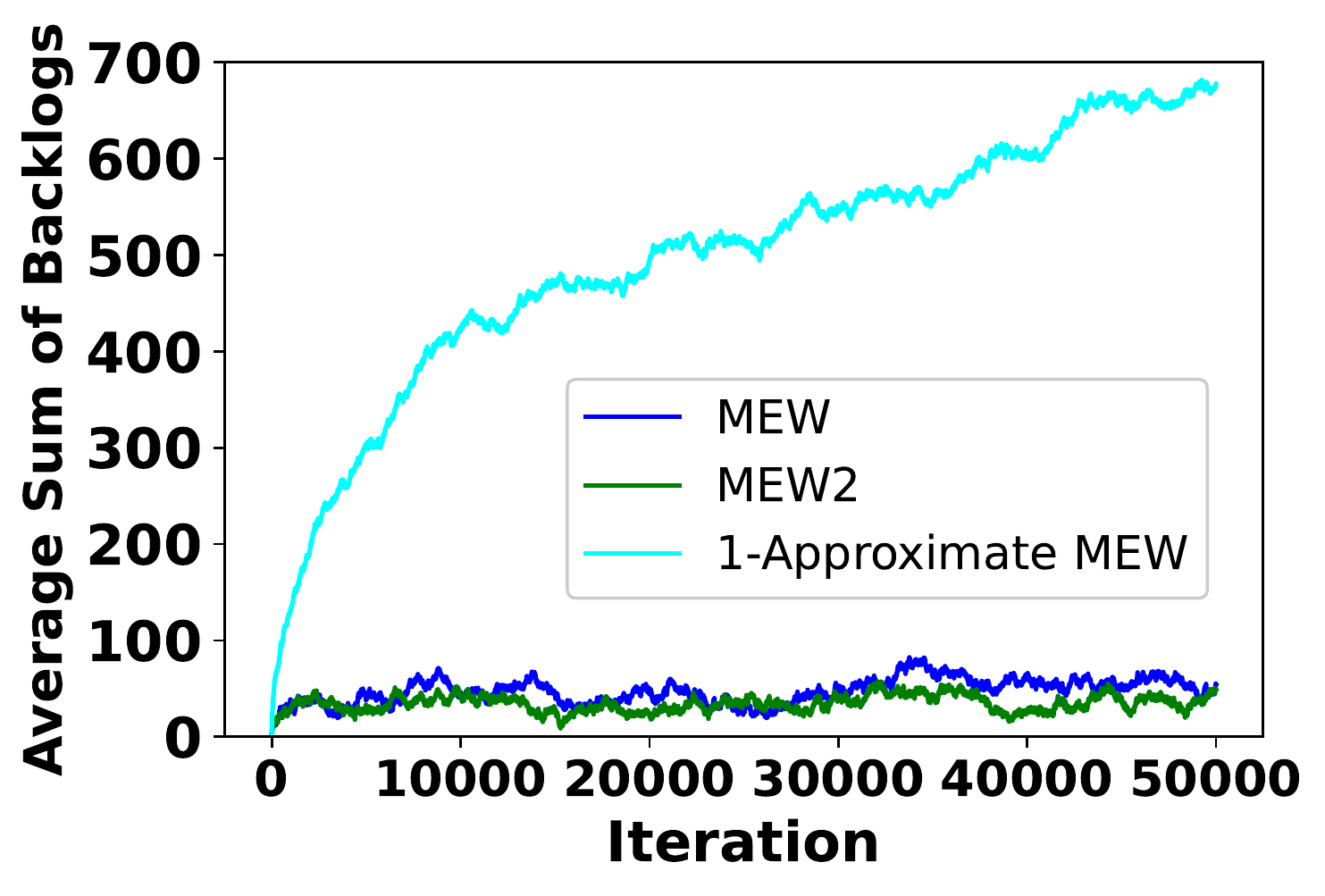}
\caption{
Illustrating the simulation in Section \ref{sec:exp3}: The evolution of MEW, MEW2 and $1$-Approximate MEW for a quantum switch with $N = 7$, $M = 4$ and bipartite requests. The traffic intensity is 99\% and the LLE attempts succeed with probability $0.9$. Each line is the average of $10$ different realizations.}
\label{fig:exp3}
\end{figure}

In this simulation, we compare MEW2 to MEW and $1$-Approximate MEW in a switch with $N = 7$ clients and $M = 4$ memories. The traffic intensity is fixed to $99\%$, and we assume that the LLE attempts are always successful. Also, all the requests for end-to-end entanglements are bipartite with $R = {7 \choose 2} =  21$. 

We run the three algorithms and show the results in Figure \ref{fig:exp3}. 
Observe from the figure that  MEW2 can keep the queues stable and that its behavior is similar to MEW. Nonetheless, recall that MEW has as higher computational cost than MEW2 (see discussion in Section \ref{sec:variant}). 
Next, observe from Figure \ref{fig:exp3} that the  $1$-Approximate MEW (which has a comparable cost to MEW2) cannot stabilize the queues.

\textbf{Conclusions:} The behavior of MEW2 is similar to MEW even though its complexity is significantly lower. The $1$-Approximate MEW does not keep the queues stable despite having a similar computational cost to MEW2.

\section{Conclusions} 
In this paper, we have studied the problem of operating a quantum switch with memory constraints. The switch has to allocate quantum memories to clients to generate link-level entanglements (LLEs), and then use these to serve end-to-end entanglements requests. The paper's main contribution is twofold: (i) to characterize the switch's capacity region, and (ii) to propose a policy (MEW) that is throughput optimal. We also present MEW2, a polynomial time variant of MEW tailored to the case where end-to-end entanglements are bipartite and LLE attempts are always successful.


\section{Appendix}

\subsection{Proof of Theorem 
\ref{th:main_theorem}}
\label{sec:appendix_proof_thm1}

We prove that the queues are stable by using a quadratic Lyapunov function, and ultimately showing that the proposed policy has expected negative drift. That is, the queues at time slots $t$ and $t+1$ satisfy: $
\mathbb E [ \| Q(t+1) \|^2 - \| Q(t) \|^2  ]  < 0$,
where the expectations are w.r.t.\ the (i) request arrivals; (ii) successful LLEs; and (iii) all possible queue values at time $t$. 
To start, observe that 
\begin{align*}
& \| Q(t+1) \|^2 = \\
& = \| [Q(t) - b(t)]^+ \|^2 + \| A(t) \|^2 + 2 \sum_{r=1}^R [ Q_r(t) - b_r(t) ]^+ A_r(t)  \\
& \le \| Q(t) - b(t) \|^2 + \| A(t) \|^2 + 2 \sum_{r=1}^R [ Q_r(t) - b_r(t) ]^+ A_r(t)  \\
& = \| Q(t) \|^2 + \| b(t) \|^2 + \| A(t) \|^2 - 2 \sum_{r=1}^R Q_r(t) b_r(t)  \\ 
& \qquad + 2 \sum_{r=1}^R [ Q_r(t) - b_r(t) ]^+ A_r (t) \\
& \le \| Q(t) \|^2 + \| b(t) \|^2 + \| A(t) \|^2 - 2 \sum_{r=1}^R Q_r(t) b_r(t)  \\
& \qquad  + 2 \sum_{r=1}^R  Q_r(t)  A_r(t)  \\
& = \| Q(t) \|^2   +  \| A(t) \|^2 + \| b(t) \|^2  + 2 \sum_{r=1}^R Q_r(t) (A_r(t) - b_r(t)).
\end{align*}

Next, since $\| A(t) \|^2 \le N^2$, $\| b(t) \|^2 \le N^2$ (by assumption), and $\mathbb E[A(t)] = \lambda$ by assumption, we can take expectations with respect to $A_r(t)$ for a fixed queue $Q(t)$ to obtain:
\begin{align}
& \mathbb E [ \| Q(t+1) \|^2 - \| Q(t) \|^2 \mid Q(t) ] \notag \\
&  \qquad  \le  2 N^2 + 2 \sum_{r=1}^R Q_r(t) (\lambda_r - b_r(t)) \label{eq:esyu}
\end{align}

We proceed to upper bound the expected value of $ - \sum_{r=1}^R Q_r(t) b_r(t)$. 
To start, because $b(t)$ is a random vector that depends on the switch connectivity and memory allocation at time $t$, we have 
\[
\mathbb E\left [ - \sum_{r=1}^R Q_r(t) b_r(t) \right ] =  - \sum_{r=1}^R Q_r(t) \mu_r(m(t),Q(t))
\]
where $\mu_r(m(t),Q(t))$ is defined in \eqref{eq:expected_service_state}. Note that $Q(t)$ does not depend on the switch connectivity in time slot $t$. Combining the last equation with \eqref{eq:esyu}, we have
\begin{equation}
\label{convcombo}
\begin{aligned}
& \mathbb E [ \| Q(t+1) \|^2 - \| Q(t) \|^2 \mid Q(t) ]  \\
& \qquad \le 2N^2 + 2 \sum_{r=1}^R Q_r(t) (\lambda_r - \mu_r(m(t),Q(t))).
\end{aligned}
\end{equation}
where the expectation is with respect to the switch connectivities for a fixed memory allocation. 

Next, observe that the memory allocation in \eqref{eq:memoryallocation} ensures that:
\begin{align}
\label{temp1}
& - \sum_{r=1}^R Q_r(t) \mu_r(m(t),Q(t)) \notag \\
& \qquad  \le - \sum_{r=1}^R Q_r(t) \mu_r({m}, Q(t)) && \forall {m} \in \mathcal{M}.
\end{align}
since $\mu(m(t), Q(t))$ maximizes $\sum_{r=1}^R Q_r(t) \mu_r(m(t),Q(t))$. 
Now, let $\theta_m \ge 0 $ with $\sum_{m\in \mathcal M} \theta_m = 1$ and observe that \begin{align*}
&  - \sum_{r=1}^R Q_r(t) \mu_r(m(t),Q(t)) \\
& \qquad =  - \sum_{m \in \mathcal M}  \theta_m \sum_{r=1}^R  Q_r(t) \mu_r(m(t),Q(t)) \\
 & \qquad  \stackrel{(a)}{\le} - \sum_{m \in \mathcal M}  \theta_m \sum_{r=1}^R  Q_r(t) \mu_r({m}, Q(t)), \\
  & \qquad  \stackrel{(b)}{=} - \sum_{m \in \mathcal M}\theta_m \sum_{r=1}^R  Q_r(t) \sum_{k \in \mathcal{K}(m)} \mathbb P(k;m) w_r(k, Q(t))  \\
    & \qquad  \stackrel{}{=} - \sum_{m \in \mathcal M} \theta_m  \sum_{k \in \mathcal{K}(m)}   \mathbb P(k;m) \sum_{r=1}^R Q_r(t) w_r(k, Q(t))  
\end{align*}
where (a) follows by \eqref{temp1} and (b)  by \eqref{eq:expected_service_state}.

Now, recall  $w(k, Q(t))$ maximizes $\sum_{r=1}^R Q_r(t) w_r(k, Q(t))$ because how we defined it in \eqref{eq:mw-update_bsm}, and let
\[
\delta^{m,k}_b \ge 0 \quad \text{for all } b \in \mathcal B(m, k),
  \quad \sum_{b \in \mathcal B(m, k)} \delta^{m,k}_b = 1
\]
for every $ k \in \mathcal K(m)$ and $m\in \mathcal M$. 
Using the same strategy as before, we have
\begin{align*}
& -  \sum_{r=1}^R Q_r(t) w_r(k, Q(t)) && k \in \mathcal K(m) \\ 
& \qquad = - \sum_{b \in \mathcal B(m, k)} \delta^{m,k}_b  \sum_{r=1}^R Q_r(t) w_r(k, Q(t)) \\
& \qquad \stackrel{(a)}{\le} - \sum_{b \in \mathcal B(m, k)} \delta^{m,k}_b  \sum_{r=1}^R Q_r(t) b_r \\
& \qquad = - \sum_{r=1}^R Q_r(t)  \sum_{b \in \mathcal B(m, k)} \delta^{m,k}_b b_r
\end{align*}
where $b$ in inequality (a) holds for any vector in $\mathcal B(m, k)$. 
Combining the previous equations, we obtain
\begin{align*}
&  - \sum_{r=1}^R Q_r(t) \mu_r(m(t),Q(t)) \le - \sum_{r=1}^R Q_r(t) \hat b_r 
\end{align*}
where
\[
\hat b = \sum_{m \in \mathcal M} \theta_m  \sum_{k \in \mathcal{K}(m)}   \mathbb P(k;m)  \sum_{b \in \mathcal B(m, k)} \delta^{m,k}_b b
\]
is any vector in $\Lambda$ (Proposition \ref{th:capacity}). Hence, from \eqref{convcombo}, we have
\begin{align*}
& \mathbb E [ \| Q(t+1) \|^2 - \| Q(t) \|^2 \mid Q(t)] \\
& \qquad   \le  2N^2+ 2 \sum_{r=1}^R Q_r(t) (\lambda_r - \hat b_r(t))
\end{align*}

The rest of the proof follows the usual max-weight arguments. Because $\lambda_r + \epsilon \le \hat b_r$ for some $\epsilon > 0$ by assumption, it holds
\begin{align*}
& \mathbb E [ \| Q(t+1) \|^2 - \| Q(t) \|^2 \mid Q(t) ] \\
& \qquad \le 2N^2 - 2 \sum_{r=1}^R Q_r(t)
 \epsilon 
\end{align*}
Now, take expectations with respect to all the possible values of $Q(t)$ to obtain
\begin{align}
& \mathbb E [ \| Q(t+1) \|^2 - \| Q(t) \|^2  ] \le 2N^2 - 2 \sum_{r=1}^R Q_r(t)\epsilon ,
 \label{eq:expecteddriftbound}
\end{align}
and observe that $ \mathbb E [ \| Q(t+1) \|^2 - \| Q(t) \|^2  ]  < 0$ when $\sum_{r=1}^R Q_r(t) > \frac{2N^2}{2\epsilon}$. That is, the queue drift is negative. 
Finally, sum \eqref{eq:expecteddriftbound} from $t=1,\dots,T$ to obtain
\begin{align*}
& \mathbb E [ \| Q(T+1) \|^2] - \mathbb E [\| Q(0) \|^2] \\
& \qquad \le 2T N^2 - 2 \epsilon  \sum_{t=1}^T \sum_{r=1}^R \mathbb E[ Q_r(t)]
\end{align*}
Rearranging terms and dividing by $T$ yields
\begin{align*}
\frac{1}{T} \sum_{t=1}^T \sum_{r=1}^R  \mathbb E[Q_r(t)]
& \le \frac{N^2}{\epsilon} + \frac{\| Q(0) \|^2}{2T\epsilon},
\end{align*}
and taking $T \to \infty$ we obtain the stated result.

\subsection{Proof of Corollary \ref{cor:1}}
\label{sec:proof_coro1}

Let $\mathcal{C}(m)$ be the complete graph that results from the clients $n \in \{1,\dots,N\}$ for which $m_n = 1$. Moreover, let $\mathcal{P}( G )$ be the set of matchings of a graph $G$, and $\mathcal{P}_N$ be the set of matchings of the complete graph with $N$ clients. The proof of Corollary \ref{cor:1} relies on finding a policy that is equivalent to MEW, which we proved in Theorem \ref{th:main_theorem} that stabilizes the queues.

We can rewrite step \textbf{(S1)} of MEW as follows:
\begin{align}
\label{simple_case_problem1}
m(t) \in \underset{{m \in \mathcal{M}}}{\arg\max} \max_{u \in \mathcal{P}(\mathcal{C}(m))} \ \sum_{r=1}^R Q_r(t) \ u_r .
\end{align}
Problem \eqref{simple_case_problem1} summarizes MEW when LLE attempts are always successful and the requests are bipartite. That holds true because (i) its solution picks the memory allocation of \textbf{(S1)}, and (ii) the maximizer of the inner optimization problem is the final service vector of step \textbf{(S3)}. Let $M^*(t)$ be the set of solutions to problem \eqref{simple_case_problem1} and define the sets

\small
\begin{align*}
 \Pi := \bigcup_{m \in \mathcal{M}} \mathcal{P}(\mathcal{C}(m)), \quad O := \{ u \in \mathcal{P}_N: \sum_{r=1}^R u_r \le M/2 \}.  
\end{align*}
\normalsize
Recall that $M$ is even. Then, step \textbf{(S3)}, is equivalent to \small
\begin{align}
\label{simple_case_b_3}
     b(t) \in & \bigcup_{m^* \in M^*(t)} \underset{{u \in  \mathcal{P}(\mathcal{C}(m^*)) }}{\arg\max}  \ \sum_{r=1}^R Q_r(t) u_r \\ 
     \nonumber
    & \stackrel{}{=} \underset{{u \in \Pi }}{\arg\max}  \ \sum_{r=1}^R Q_r(t) u_r \\
    \label{simple_case_b_2}
    & \stackrel{}{=} \underset{{u \in O }}{\arg\max}  \ \sum_{r=1}^R Q_r(t) u_r.
\end{align}
\normalsize
Therefore, step \textbf{(S3)} of MEW can be solved immediately by solving the problem \eqref{simple_case_b_2} (see step (\textbf{S1b}) in MEW2). 
Note that \eqref{simple_case_b_2} does not depend on the optimal set $M^{*}$ and therefore we can solve it before allocating the quantum memory. However, to characterize the policy we have to find a memory allocation that would make $b(t)$ computed in \eqref{simple_case_b_2} a feasible service vector. In \eqref{simple_case_problem1} note that any memory allocation that includes the clients involved in the matching $b(t)$ (step (\textbf{S1b})), belongs in $M^{*}$ and therefore would be an optimal memory allocation.


\bibliographystyle{IEEEtran}
\bibliography{IEEEabrv, references}

\end{document}